\documentclass[prl,twocolumn,showpacs,floatfix]{revtex4}
\usepackage{graphicx}
\usepackage{dcolumn}
\usepackage{amsfonts}
\usepackage{bm}

\begin{document}
\title{Two-electron quantum dot molecule:\\ Composite particles and
the spin phase diagram} 
\author{A. Harju} 
\author{S. Siljam\"aki}
\author{R.M. Nieminen}

\affiliation{Laboratory of Physics, Helsinki University of Technology,
P.O. Box 1100, 02015 HUT, Finland}

\date{\today}
\begin{abstract}
We study a two-electron quantum dot molecule in a magnetic field by
the direct diagonalization of the Hamiltonian matrix.  The ground
states of the molecule with the total spin $S=0$ and $S=1$ provide a
possible realization for a qubit of a quantum computer. Switching
between the states is best achieved by changing the magnetic field.
Based on an analysis of the wave function, we show that the system
consists of composite particles formed by an electron and flux quanta
attached to it.  This picture can also be used to explain the spin
phase diagram.
\end{abstract}


\maketitle

The progress in semiconductor technology has opened a rich field of
studies focused on the fundamental electron-electron interactions and
quantum effects in artificial atoms and molecules~\cite{qdrev}.  The
most striking feature of two-dimensional semiconductor quantum dots
(QD) and quantum molecules (QDM) is that the correlation and magnetic
field effects are greatly enhanced compared with their normal
counterparts. This results in a rich variety of phenomena in lateral
QDMs that have recently been investigated experimentally and
theoretically, see for example Refs.~\cite{eqdm1,eqdm2,wen,tqdm}.
Also the system parameters can easily be changed, unlike in real atoms
and molecules where the parameters are natural constants.  The
controllable parameters make it possible to tailor the semiconductor
structures and for example to switch between different ground states.

In this Letter we concentrate on a two-electron QDM consisting of two
laterally coupled QDs.  In addition to the interesting and fundamental
correlation and quantum effects, this system is very important as a
candidate for the gate of a quantum computer~\cite{qc}. A central idea
is to use the total spin ($S$) of the two-electron QDM as a qubit.
The ground state spin of the QDM can be either $S=0$ or $S=1$.  One of
the aims of this Letter is to study the regions of $S=0$ and $S=1$
states as a function of the important system parameters in their
realistic range, beyond the approximations of Ref.~\onlinecite{qc}
where a change in $S$ was found when the magnetic field $B$ was
varied.  We find a similar crossing at weak $B$, but also a
reappearance of the $S=0$ ground state at larger $B$. This is caused
by the correlation effects that are treated accurately in our
many-body approach. An interesting finding is that the system
naturally consists of composite particles of electrons and flux quanta
as in the composite-fermion theory~\cite{cf}.

\begin{figure}[htb]
 \includegraphics[height=4.7cm]{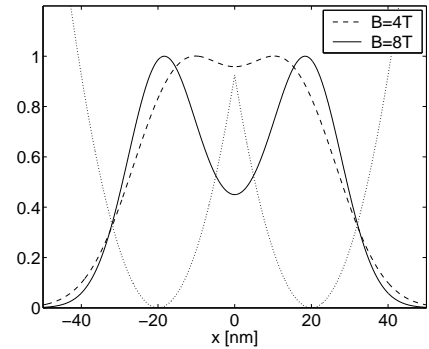}
 \caption{Confinement potential $V_\mathrm{c}$ (dotted line) and
 non-interacting single-particle density $|\psi|^2$ along $y=0$ for
 $d=40$~nm.  The potential is in units of $\hbar \omega_0=3$~meV, and
 the maximum value of each $|\psi|^2$ is scaled to unity.  The
 confinement potential is parabolic in the $y$-direction.  One can see
 a localizing effect of large $d$ and strong $B$.}
\label{fig1}
\end{figure}

\begin{figure}[htb]
 \includegraphics[height=4.7cm]{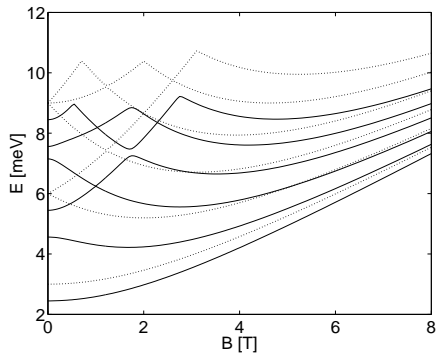}
 \caption{Six lowest non-interacting single-particle energies as a
 function of $B$.  The dotted lines are the normal Fock-Darwin states
 ($d=0$), and the solid ones are for $d=20$~nm.  One can see that for
 $d>0$, the energies are shifted down and that the level crossings
 occur at weaker $B$.  At some of the level-crossings a gap opens,
 suchs as the one for $B\approx 1.75$~T, $E\approx 7.4$~meV.  This is
 due to the lower symmetry of the problem for $d>0$.  In general,
 there is a reasonable similarity between the two sets of energy
 levels.  For sufficiently large $d$, the energies would converge to
 degenerate states of isolated dots.  All states are on the lowest
 Landau level after $B\approx 3$~T.}
\label{fig2}
\end{figure}

We model the two-electron QDM by a 2D Hamiltonian
\begin{equation}
H = \sum _{i=1}^2\left ( \frac{ ( {-\mathrm{i}{\hbar} \nabla_i}
-\frac ec \mathbf{A})^2 }{2 m^{*}} + V_\mathrm{c}({\bf r}_{i}) \right ) +
\frac {e^{2}}{\epsilon r_{12}} ,
\label{ham}
\end{equation}
where $V_\mathrm{c}$ is the external confining potential, for which we
use the one of Ref.~\cite{wen}, namely $\frac 12 m^* \omega_0^2 \min
\{(x-d/2)^2+y^2, (x+d/2)^2+y^2\}$.  This
potential separates to two QDs at large inter-dot distances $d$, and
with $d=0$ it simplifies to one parabolic QD.  We use GaAs material
parameters $m^*/m_e=0.067$ and $\epsilon=12.4$, and the confinement
strength $\hbar\omega_0=3.0$~meV.  $\mathbf{A}$ is the vector
potential of the magnetic field (along the $z$ axis) taken in the
symmetric gauge.  One should note that the Hamiltonian is spin-free,
and the Zeeman-coupling $E_{Z}=g^* \mu_B B S_z$ (with $g^* = -0.44$
for GaAs) of the magnetic field to $S_\mathrm{z}$ can be taken into
account afterwards.  The eigenstates of the single-particle part of
Eq.~(\ref{ham}) are easily obtained as expansions
\begin{equation}
\psi(\mathbf{r})= \sum_i \alpha_i \phi_i(\mathbf{r}) =\sum_i
\alpha_i x^{n_{x,i}} y^{n_{y,i}} e^{-\frac 12 r^2} \ ,
\end{equation}
where $n_x$ and $n_y$ are integers, and $\alpha_i$ is a complex
coefficient.  We have used the unit of length as in Ref.~\cite{qd_cl}.
Fig.~\ref{fig1} shows examples of the non-interacting charge
densities, and Fig.~\ref{fig2} displays single-particle energies.
It is interesting to compare the localizing effect of $B$ with the
experimental findings of Brodsky \textit{et al.}~\cite{eqdm2}. They
see a clear splitting of the QDM electron droplet to smaller fragments
by a strong $B$. As they work in a low-density limit (weak
confinement), impurities can have a similar role as the potential
minima in Fig.~\ref{fig1}.  The localization is also related to the
formation of Wigner molecules in QDs~\cite{qd_cl}, which happens in
the low-density limit.

Similarly to the single-particle states, the full many-body wave
function with total spin $S$ can be expanded as
\begin{eqnarray}
\Psi_S (\mathbf{r}_1, \mathbf{r}_2) &=& \sum_{i\le j} \alpha_{i,j}
\left\{ \phi_i(\mathbf{r}_1) \phi_j(\mathbf{r}_2) \right.\nonumber \\
&+& \left.(-1)^S 
\phi_i(\mathbf{r}_2) \phi_j(\mathbf{r}_1) \right\}
\end{eqnarray}
which is symmetric for $S=0$ and anti-symmetric for $S=1$.  Notice
that the spin part of the wave function is not explicitly written, and
we work with spin-independent wave functions also below. The
coefficient vector ${\bm \alpha}_l$ and the corresponding energy $E_l$
for the $l$th eigenstate are found from a generalized eigenvalue
problem where the Hamiltonian and overlap matrix elements can be
calculated analytically. Details of the computational procedure will
be published elsewhere~\cite{longer}.  As the basis functions we have
used all the states with both $n_{x}$ and $n_{y}\le n$ with $n=6$, and
we have checked the convergence by varying $n$ for all values of d.

\begin{figure}[htb]
 \includegraphics[width=\columnwidth]{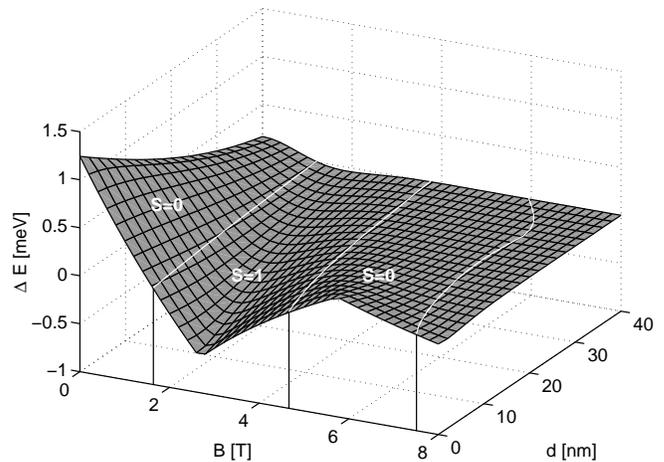}
 \caption{Energy difference between singlet and triplet states $\Delta
 E$ as a function of the magnetic field $B$ and inter-dot separation
 $d$.  The white lines separate the $S=0$ ($\Delta E>0$) and $S=1$
 ground states.}
\label{qdm_dE}
\end{figure}

\begin{figure}[htb]
 \includegraphics[width=.8\columnwidth]{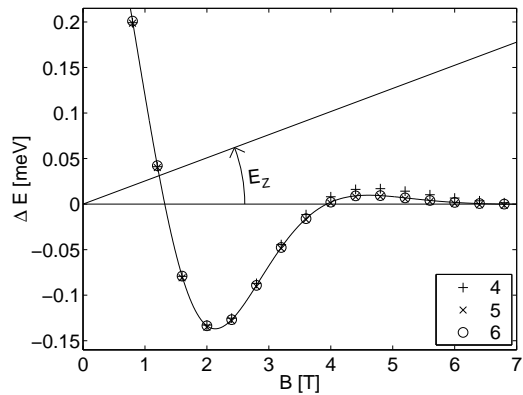}
 \caption{$\Delta E$ for a fixed $d=26.736$~nm~\cite{henri}.  The
 effect of the Zeeman energy $E_Z$ is also shown.  The difference
 between expansions with $n=5$ and 6 is only around 1~$\mu$eV.  One
 can see that the second $S=0$ state disappears for even a weak
 $E_Z$.}
\label{qdm_dE1}
\end{figure}

The energy difference $\Delta E$ between the lowest $S=0$ and $S=1$
states is plotted in Figs.~\ref{qdm_dE} and \ref{qdm_dE1}.  The
convergence of $\Delta E$ can be seen in Fig.~\ref{qdm_dE1} for large
$d$.
One can see that for weak magnetic field values, the ground state has
$S=0$.  We first concentrate on this regime.  For $B=0$ the $S=0$
state remains lower with arbitrary $d$, see e.g. Ref.~\cite{wen} for a
discussion of this exact property.  However, there is a strong
decrease in $\Delta E$ as a function of $d$.  This can be understood
from the fact that very distant QDs interact only weakly and the
energy difference of the two spin states is smaller. There is also a
strong decrease in $\Delta E$ as a function of $B$ in the $S=0$ state.
For $d=0$, $\Delta E$ is rather linear until the crossing to the $S=1$
state, but for $d > 0$ the curves are rounded.  The crossing point of
the different ground states does not depend strongly on $d$; it
changes only from 1.6~T to 1.2~T as one moves from $d=0$ to 40~nm.
Due to this, changing the total spin in an experimental setup is not
easy by just changing $d$.  On the other hand, around $\Delta E=0$,
the slope of $\Delta E$ is rather large and changing $S$ by $B$ is the
most natural choice.  One can achieve a change of $S$ also by changing
the strength of $V_\mathrm{c}$. This changes the ratio between the
energies resulting from the confinement and electron-electron
interaction.  For a weak $V_\mathrm{c}$, the interactions are stronger
and the transition occurs at weaker $B$ value.  Thus the change of the
$V_\mathrm{c}$ can be seen as a change of the effective value of $B$.

The transition from the weak-$B$ $S=0$ state to the $S=1$ state is
most simply explained by the fact that the energies of the two lowest
single-particle states approach each other as $B$ is made stronger
(see Fig.~\ref{fig2}).  At some point this difference is smaller
than the exchange-energy, and the system spin-polarizes.  One can see
that also for this state, $\Delta E$ decreases strongly as a function
of $d$.

\begin{figure}[tbh]
 \includegraphics[width=5cm]{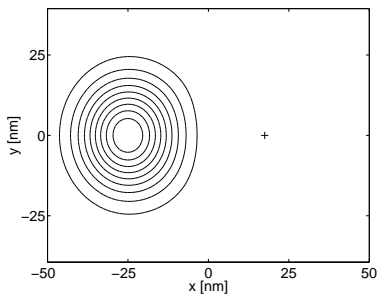}
 \caption{ $|\Psi[(x,y),(d/2,0)]|^2$ for the state $S=0$ at $B=4.5$~T
 and $d=35$~nm. The contour spacing is uniform. We mark with a plus
 the fixed electron on the right-hand QD. One can see that the
 electron density is localized to the left QD.  There is only a small
 deformation from circular symmetry.  Notice that the effective
 inter-dot distance is much larger than $d$ due to the Coulomb
 repulsion of the electrons.}
 \label{qdm_rho} 
\end{figure}

\begin{figure}[tbh]
 \includegraphics[width=8.5cm]{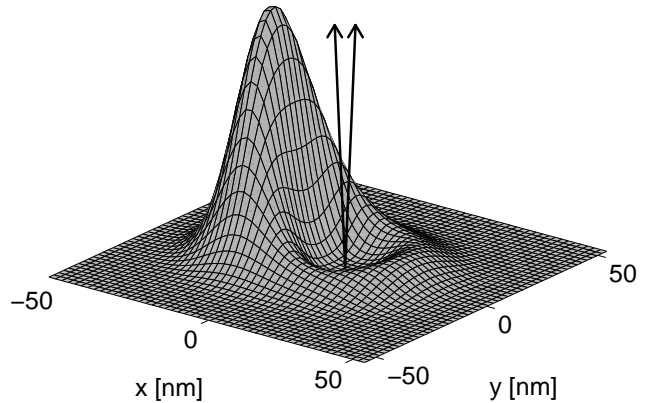} 
 \caption{ $|\Psi[(x,y),(d/2,0)]|^2$ for the state $S=0$ at $B=6$~T
 and $d=10$~nm.  The arrows (rotated for clarity) depict the two flux
 quanta at the electron fixed at $x>0$ minimum of $V_\mathrm{c}$.}
 \label{cf_rho} 
\end{figure}

There exists a second region of $S=0$ ground state around $B\approx
6$~T at $d=0$.  The question whether this state terminates at large
$d$ remains open, as the energy differences at $d>40$~nm are smaller
than the error made in the expansion.  The existence of this $S=0$
state for small value of $d$ can be understood on the basis of a
parabolic two-electron QD, which can be shown to have the exact wave
function of the form $\Psi = (x_{12} + \mathrm{i} y_{12} )^m f(r_{12})
e^{-\frac 12 (r_1^2+r_2^2)}$, where $m$ is the relative angular
momentum and $f$ is a correlation factor~\cite{2eQD}.  The simple form
is due to the separation of the center-of-mass and relative motion of
the electrons.  For $B=0$ the ground state has $m=0$ (and $S=0$), and
when $B$ is made stronger, the ground state $m$ has increasing
positive integer values.  These transitions happen because the larger
$m$ states have smaller Coulomb repulsion between the electrons, and
as the single-particle energies group together to form the lowest
Landau level, it is favorable to move to larger $m$.  The second $S=0$
region in Fig.~\ref{qdm_dE} corresponds to $m=2$.  It is surprising
that the second $S=0$ region extends to such a large $d$.  For
example, at $B=4.5$~T and $d=35$~nm the two QDs of a QDM are rather
decoupled (see Fig.~\ref{qdm_rho}), but still the $S=0$ state remains
the ground state.
To analyze the structure of this state for small and large values of
$d$, we have located the vortices of $\Psi$.  This can be done by
finding the zeros of $\Psi$ and studying the change in the phase of
$\Psi$ in going around each of the zeros.  A surprising finding is
that for both large $d$ and $d=0$ there are two vortices at both
electron locations, see Fig.~\ref{cf_rho} for an example.  The
particles are thus composite particles of an electron and two flux
quanta, in similar fashion as in the composite fermion (CF)
theory~\cite{jain95EPL}.  The most remarkable feature of the finding
is the stability of composite particles against the change of $d$.

We have done a similar analysis for the $S=1$ state at $B=3$~T and
various values of $d$.  We found that the many-body state again
consists of composite particles, but this time there is only one flux
quantum per electron.  One should note that for an odd number of flux
quanta per electron, the wave function changes sign when the particles
are exchanged, corresponding to $S=1$.  Similarly, for even flux
numbers the state has $S=0$.

If one expands $|\Psi|^2$ for small
$r_{12}$, one obtains
\begin{equation}
|\Psi|^2 \propto r_{12}^{2 m}+ \frac{C}{m + \frac 12}r_{12}^{2 m + 1}
 + \mathcal{O}(r_{12}^{2m+2})\ ,
\end{equation}
where $C$ is the scaled strength of the Coulomb
interaction~\cite{qd_cl}, and $m\ge 0$ is the number of flux quanta
per electron.  One can see that for larger $m$ values, the density
grows more slowly as a function of $r_{12}$ (see Fig.~\ref{cf_rho}).
One should note that the same expansion is valid also for the larger
electron numbers, resulting from the cusp condition~\cite{qd_cl}. Thus
the electron-electron interaction is smaller when the number of flux
quanta per electron grows.

One can use the CF-type approach to explain the phase diagram of
Fig.~\ref{qdm_dE}.  When one moves from $B=0$ to stronger values of
$B$, the number of flux quanta in the system increases.  At the first
$S=0$ state there is no fluxes in the system, and at the transition
point the flux number changes to one per electron.  For weakly coupled
QDs this transition happens at smaller $B$ than in the strongly
coupled ones, because a distant zero of the wave function increases
the kinetic energy less than a close one does.  In the following
transition points the number of flux quanta changes again by one per
electron.  The reasoning for the $d$-dependence of the first
transition applies to other ones also, and this can be seen in
Fig.~\ref{qdm_dE}.

An interesting prospect resulting from the discussion above is to use
the CF-type approach to describe the many-body states of electrons at
strong $B$ in various confining potentials.
One should note that after the phase structure of the wave function is
fixed, one is left with the bosonic part of the wave function.  The
quantum Monte Carlo techniques are especially useful for obtaining
this part~\cite{fpqmc}.

If one adds the Zeeman term to the Hamiltonian, the energy of the
$S=1$ state is lowered by $\sim 25$~$\mu$eV/T and the $S=0$ energy is
unaltered, see Fig.~\ref{qdm_dE1}.  This makes the high-$B$ $S=0$
state to terminate at $d$ value around 5~nm.  One should note that it
is possible to lower the Zeeman term also in the experimental setup by
applying a tilted magnetic field.  Most probably the singlet-triplet
separation in energy for $B\approx 6$~T is too small for a qubit, but
in principle it is possible to change $S$ by varying the Zeeman term
in this parameter range.

In conclusion, we have determined the total-spin phase diagram of the
two-electron quantum dot molecule as a function of the magnetic field
and the inter-dot distance.  Our results support the possibility to
use the system for a gate of a quantum computer~\cite{qc}.  In
addition, we have found that the system consists of composite
particles of electrons and the attached magnetic field flux quanta.

\begin{acknowledgments}
This research has been supported by the Academy of Finland through its
Centers of Excellence Program (2000-2005).
\end{acknowledgments}

\end{document}